\documentclass[12pt]{article}
0.1cm \evensidemargin - 0.5cm
\newcommand{\eproof}{\rule{0.2cm}{0.2cm}}
\newcommand{\re}{I\!\!R}
\newcommand{\R}{I\!\!R}

\newcommand{\ds }{ \displaystyle }
\newcommand{\nat}{I\!\!N}
\newcommand{\vf}{\varphi}

\usepackage{epsfig}
\usepackage{amssymb,amsmath}

\newtheorem{theorem}{Theorem}[section]
\newtheorem{lemma}{Lemma}[section]
\newtheorem{corollary}{Corollary}[section]
\newtheorem{definition}{Definition}[section]
\newtheorem{remark}{Remark}[section]

\begin{document}
\title{\bf Variable order differential equations and diffusion
processes with changing modes}
\author{Sabir Umarov{$^{\ast}$}, Stanly Steinberg{$^\dagger$}}
\date{}
\maketitle
\begin{center}
{ $^{*}$ Department of Mathematics, Tufts University, Medford, MA,
USA\\
$^{\dagger}$ Mathematics and Statistics Department, University of
New Mexico, Albuquerque, NM, USA}
\end{center}

\begin{abstract}
% Text of abstract
In this paper diffusion processes with changing modes are studied
involving the variable order partial differential equations. We
prove the existence and uniqueness theorem of a solution of the
Cauchy problem for fractional variable order (with respect to the
time derivative) pseudo-differential equations. Depending on the
parameters of variable order derivatives short or long range
memories may appear when diffusion modes change. These memory
effects are classified and studied in detail. Processes that have
distinctive regimes of different types of diffusion depending on
time are ubiquitous in the nature. Examples include diffusion in a
heterogeneous media and protein movement in cell
biology.
\end{abstract}


\section{Introduction}
Diffusion processes can be classified according to the asymptotic
behaviour of their mean square displacement (MSD) as a function of
time. If the dependence of the MSD on time is linear then the
process is classified as normal, otherwise as anomalous. For many
processes, the MSD satisfies
\begin{equation}
\label{MSD1} \mbox{MSD}(t) \sim K_{\beta} t^{\beta} \,,\quad t
\rightarrow \infty,
\end{equation}
where $K_{\beta}$ is a constant.  If $\beta = 1$ the diffusion is
normal, if $\beta >1$ the process is super-diffusive, while if
$\beta < 1$ the process is sub-diffusive \cite{MK,Zaslavsky}.  The
ultra-slow diffusion processes studied in \cite{Chechkin1,MS,MK2004}
lead to logarithmic behaviour of MSD for large $t$. The MSD of more
complex processes with retardation (see
\cite{Chechkin1,GorenfloMainardi2005}) behaves like $t^{\beta_2}$
for $t$ small, and $t^{\beta_1}$ for $t$ large, where $\beta_1 <
\beta_2$. Subdiffusive motion with $0.1<\beta<0.9$ was recorded in
\cite{GhoshWebb}, and with $0.22 < \beta < 0.48$ in
\cite{Slattery95}, depending on macromolecules and cells. In
\cite{Kusumi2, Kusumi} protein movement is studied in the cell
membrane with a few types of compartments and made a conclusion that
$\beta$ depends on time scales. Our models are subdiffusive, but of
variable order with {\em order function} $\beta = \beta(t)$.

It is well known that simple homogeneous subdiffusive processes can
be modeled using a fractional order partial differential equation
where only the time derivative has a constant fractional order
\cite{MK}.  Variable fractional order derivatives and operators were
studied by N. Jacob, at al. \cite{Jacob93}, S. G. Samko, at al.
\cite{Samko1, Samko2}, W. Hoh \cite{Hoh2000}. Recently A. V.
Chechkin, at al. \cite{Chechkin4} used a version of variable order
derivatives to describe kinetic diffusion in heterogeneous media. In
the recent paper \cite{LorenzoHartley}, Lorenzo and Hartley
introduced several types of fractional variable order derivatives
and applied them to engineering problems. We will modify these
operators, restrict them to order functions $\beta(t)$ that are
piecewise constant and then apply the resulting variable order
partial differential equations (VOPDE) to diffusion processes with
changing diffusion modes. An important aspect of the modeling is
that the VOPDEs provides a description of memory effects arising
from a change of diffusion modes that are distinct from the ``long
range memory'' connected with the non-Markovian character of
diffusion.  Thus, in the VOPDE based description of anomalous
diffusion models, both non-Markovian long range memory and new type
of memory may be present simultaneously.

The paper is organized as follows. In Section 2 we introduce
background material. In Section 3 we study the  memory effects
arising in connection with a change of diffusion modes. In Sections
4 and 5 we study the mathematical model of diffusion processes with
changing modes in terms of an initial value problem for VOPDE.
Namely, we prove the theorem on the existence and uniqueness of a
solution of the initial value problems for variable order
differential equations and study some properties of a solution.
%Note that
The theorems are proved under the assumption that diffusion mode
change times are known.

%The general case with random change of modes $\beta$ and random mode
%change times $T$ will be treated in a separate paper, as well as an
%asymptotic behaviour of MSD for large $t$ in the general case. The
%latter requires a delicate analysis of the symbol of the solution
%operator.

\vspace{1cm}

\section{LH-parallelogram and variable order derivatives}
Recently  Lorenzo and Hartley \cite{LorenzoHartley} introduced three
types of derivatives of variable fractional order $\beta(t)$, $t>0$,
$0 < \beta(t) \leq 1$ all of which are special case of a more
general fractional order derivative
\begin{equation}
\label{RLLH} \mathcal{D}_{\mu,\nu}^{\beta(t)} f(t) = \frac{d}{dt}
\int_0^t \mathcal{K}_{\mu,\nu}^{\beta(t)}(t,\tau)f(\tau)d\tau,
\end{equation}
where $\mu$ and $\nu$ are real parameters, $t>0$, and
\begin{equation}
\label{kernel} \mathcal{K}_{\mu,\nu}^{\beta(t)}(t, \tau)=
\frac{1}{\Gamma(1-\beta(\mu t+\nu \tau))(t-\tau)^{\beta(\mu t+\nu
\tau)}}, \quad 0<\tau < t \,.
\end{equation}
For convenience in studying of initial value problems, we prefer to
use the closely related Caputo type operator
\begin{equation}
\label{CaputoLH} \mathcal{D}_{\ast \mu,\nu}^{\beta(t)} f(t) =
\int_0^t \mathcal{K}_{\mu,\nu}^{\beta(t)}(t,\tau)\frac{df(\tau)}{d
\tau} d\tau.
\end{equation}

To describe the properties of the kernel (\ref{kernel}) and the
fractional derivative operators \eqref{RLLH} and \eqref{CaputoLH} we
introduce the Lorenzo-Hartley (LH) {\em causality parallelogram}
\cite{LorenzoHartley} $ \Pi = \left\{(\mu,\nu)\in R^2 : 0 \leq \mu
\leq 1, -1 \leq \nu \leq +1, 0 \leq \mu+\nu \leq 1 \right\}. $ The
kernel \eqref{kernel}, and thus, both the operators \eqref{RLLH} and
\eqref{CaputoLH} are weakly singular for $(\mu, \nu) \in \Pi$.
Further, denote
\begin{equation}
\label{kis} \mathcal{K}(t, \tau, s)= \frac{1}{\Gamma(1-\beta
(s))(t-\tau)^{\beta(s)}}, \, t>0, \, 0<\tau < t, \, s \geq 0,
\end{equation}
where $0 < \beta(s) \leq 1$
%alteration
\footnote{if $\beta(t)=1,$ then we agree $\mathcal{D}_{\ast
\mu,\nu}^{\beta(t)}f(t)=\frac{d f(t)}{d t}.$} is a given function,
which is called an {\it order function}.

Our main goal is to model problems where for different time
intervals there are different
% this is alteration
modes\footnote{see definition in Section \ref{subdiffusion}.} of
diffusion. To this end, let $T_i$ be a partition of the interval
$(0,\infty)$ into $N+1$ sub-intervals $(T_k,T_{k+1})$, where
$0=T_0<T_1<...<T_N < T_{N+1} = \infty$.  Then let $\beta(t)$ be the
piecewise constant function
\begin{equation}
\label{beta} \beta(t)=\sum_{k=0}^{N} \beta_k \,
\mathcal{I}_k(t)\,,\quad t\in(0,\infty),
\end{equation}
where $\mathcal{I}_k$ is the indicator of the interval $(T_{k},
T_{k+1})$ and $0 < \beta_k \leq 1, \, k=0,...,N,$ are constants.
Under these conditions, the function \eqref{kis} becomes
% alteration
\begin{equation}
\label{kist} {K}(t, \tau, s)= \sum_{k=0}^N \mathcal{I}_k (s)
\frac{1}{\Gamma(1-\beta_k)(t-\tau)^{\beta_k}}, \, t>0, \, 0<\tau <
t, \, s \ge 0,
\end{equation}
and the kernel of the fractional order operator \eqref{CaputoLH}
becomes
% alteration
\begin{equation}
\label{KERNEL} \mathcal{K}_{\mu,\nu}^{\beta(t)} (t, \tau) = K(t,
\tau,  \mu t + \nu \tau), \, t > 0, \, 0 \le \tau < t.
\end{equation}
with $K(t,\tau,s)$ defined in (\ref{kist}).
% alteration (eliminated)
%\begin{equation}
%\label{kernelpq} K_{\mu,\nu}^{\beta(t)}(t, \tau, s) =
%\mathcal{K}_k(t, \tau, s), \, t>0, \, 0<\tau < t, \, s \geq 0.
%\end{equation}

We think of the input to our model as the triplet $(\beta_k, \mu,
\nu)$, $0 \leq k \leq N,$ while the output of our model is
determined by the kernel \eqref{KERNEL}.  Correspondingly, we say
that the triplet $(\beta_k, \mu, \nu)$ determines the diffusion mode
in the time interval $(T_{k},T_{k+1}).$ The output is determined by
which values of $\beta(t)$ are used to compute the variable order
derivative, that is, by which interval $(T_{k},T_{k+1})$ the point
$\mu t+\nu \tau$ belongs to. We always assume that $(\mu,\nu) \in
\Pi$ and then note that $\tau \in (0,t)$ yields $\mu t+\nu \tau \in
(\mu t, (\mu+\nu)t)$ and that $(\mu t, (\mu+\nu)t) \subset (0,t)$.
This means that the operators $\mathcal{D}_{\mu,\nu}^{\beta(t)}$ and
$\mathcal{D}_{\ast \mu,\nu}^{\beta(t)}$ use information taken in the
time sub-interval $(\mu t, (\mu+\nu)t)$ if $\nu$ is positive and
from the sub-interval $ ((\mu+\nu)t, \mu t)$ if $\nu$ is negative.
In both cases, the length of this interval is $| \nu | \, t$. The
condition $(\mu, \nu) \in \Pi$ predetermines the causality, since $0
\leq \mu t +\nu \tau \leq t$ for all $t > 0$ and $0  \leq \tau \leq
t.$

%alteration (typos are corrected)
\subsection{Generalized function spaces $\Psi_{G,p}(\R^n)$, $\Psi^{'}_{-G,q}(\R^n)$ }

Let $p>1$, $q>1$, $p^{-1}+q^{-1}=1$ be two conjugate numbers. The
generalized functions space $\Psi_{-G,q}(\R^n)$, which we are going
to introduce is distinct from the classical spaces of generalized
functions. In the particular case of $p=2$ this space was first used
by Yu.A.Dubinskii \cite{Dubinskii} in the course of initial-value
problems for pseudo-differential equations with analytic symbols.
Later, the general case for $\forall \, p $ was studied in
\cite{Umarov97, Umarov98}. Here we briefly recall some basic facts
related to these spaces, referring the interested reader to
\cite{Umarov97, GLU} for details.

Let $G\subset \R^n$ be an open domain and a system $\mathcal{G}
\equiv \{g_k\}_{k=0}^\infty$ of open sets be a locally finite
covering of $G$, i.e., $G=\bigcup_{k=0}^{\infty}g_k$, $g_k \subset
\subset G$. This means that any compact set $K\subset  G$ has a
nonempty intersection with a finite number of sets $g_k$. Denote by
$\{\phi_k\}_{k=0}^{\infty}$ a smooth partition of unity for $G$.
We set $G_N=\cup_{k=1}^{N}g_k$ and
$\kappa_N(\xi)=\sum_{k=1}^{N}\phi_k(\xi).$ It is clear that
$G_N\subset G_{N+1},\  N=1,2,\dots,$ and $G_N\rightarrow G$ for
$N\rightarrow\infty$. Further, by $Ff$ (or $\hat{f}(\xi)$) for a
given function $f(x)$ we denote its Fourier transform, formally
setting
% alt (formal definition of Fourier transform introduced)
$Ff(\xi)=\int_{\R^n} f(x)e^{ix\xi}dx,$ and by $F^{-1}\hat{f}$ the
inverse Fourier transform, i.e.
%alt
$F^{-1}\hat{f}(x)=(2\pi)^{-n}\int_{\R^n} \hat{f}(\xi)e^{-ix\xi}d
\xi.$ The support of a given $f$ we denote by $\mbox{supp} f$.

\begin{definition}
Let $N \in \nat$. Denote by $\Psi_{N,p}$ the set of functions $f\in
L_{p}(\R^{n})$ satisfying the conditions (1)--(3):
\begin{enumerate}
\item[(1)]
$\mbox{supp } Ff\subset G_N$;
\item[(2)]
$\mbox{supp } Ff \cap \mbox{supp }\phi_{j} =\emptyset$ for $j>N$;
\item[(3)]
$p_N(f)=\|F^{-1}\kappa_{N}Ff\|_p<\infty$.
\end{enumerate}
\end{definition}

\begin{lemma}
\label{lemma2} For $N=1,2,\dots,$ the relations
\[\Psi_{N,p} \hookrightarrow
\Psi_{N+1,p} \,, \, \, \, \, \, \, \, \Psi_{N,p} \hookrightarrow
L_p(\R^n)\] hold, where $\, \hookrightarrow \, $ denote the
operation of continuous embedding.
\end{lemma}

It follows from Lemma \ref{lemma2} that $\Psi_{N,p}$ form an
increasing sequence of Banach spaces. Its limit with the inductive
topology we denote by $\Psi_{G,p}.$
%Definition of the space

\begin{definition}
$\Psi_{G,p}(\R^n)=   \mbox{{\rm
ind}}\,\lim_{N\rightarrow\infty}\Psi_{N,p}.$
\end{definition}

The inductive limit topology of $\Psi_{G,p}(\R^n)$ is equivalent to
the following convergence.

\begin{definition}
A sequence of functions $f_m \in\Psi_{G,p}(\R^n)$ is said to
converge to an element $f_0 \in\Psi_{G,p}(\R^n)$ iff:
\begin{enumerate}
\item
there exists a compact set $K\subset  G$ such that $\mbox{supp}\,
\hat{f}_m\subset K $ for all ${m\in \nat}$;
\item
$\|f_m-f_0\|_p=(\int_{R^n}|f_m-f_0|^p dx)^\frac{1}{p}\rightarrow 0$
for $m\rightarrow \infty$.
\end{enumerate}
\end{definition}

\begin{remark}
According to the Paley-Wiener-Schwartz theorem, elements of
$\Psi_{G,p}(\R^n)$ are entire functions of exponential type which,
restricted to $\R^n$, are in the space $L_p(\R^n)$.
\end{remark}

The space topologically dual to $\Psi_{G,p}(\R^n)$, which is the
projective limit of the sequence of spaces conjugate to
$\Psi_{N,p}$, is denoted by  $\Psi_{-G,q}^{'}(\R^n).$

\begin{definition}
$\Psi_{-G,q}^{'}(\R^n) =  \mbox{{\rm
pr}}\,\lim_{N\rightarrow\infty}\Psi_{N,p}^{\ast}.$
\end{definition}

In other words,  $\Psi_{-G,q}^{'}(\re^n)$ is the space of all linear
bounded functionals defined on the space $\Psi_{G,p}(\R^n)$ endowed
with the weak topology. Namely, a sequence of generalized functions
$g_N\in\Psi_{-G,q}^{'}(\R^n)$ converges to an element
$g_0\in\Psi_{-G,q}^{'}(\re^n)$ in the weak sense if for all
$f\in\Psi_{G,p}(\R^n)$ the sequence of numbers $<g_N,f>$ converges
to $<g_0,f>$ as $N\rightarrow\infty$. We recall that the notation
$<g,f>$ means the value of $g \in \Psi_{-G,q}^{'}(\R^n)$ on an
element $f \in \Psi_{G,p}(\R^n)$.

\subsection{Pseudo-differential operators with constant symbols.}
Now we recall some properties of pseudo-differential operators with
symbols defined and continuous in a domain $G\subset \re^n$. Outside
of $G$ or on its boundary the symbol $A(\xi)$ may have singularities
of arbitrary type. For a function $\varphi \in \Psi_{G,p}(\re^n)$
the operator $A(D)$ corresponding to $A(\xi)$ is defined  by the
formula
%(PDO)
\begin{equation}
\label{4} A(D)\varphi (x)= \frac{1}{(2 \pi)^n} \int_G A(\xi) F
\varphi (\xi) e^{i x \xi} d \xi.
\end{equation}
Generally speaking, $A(D)$ does not make sense even  for functions
in the space $C_0^{\infty}(\re^n)$. In fact, let $\xi_0$ be a
non-integrable singular point of $A(\xi)$ and denote by $ O(\xi_0)$
some neighborhood of $\xi_0$. Let us take a function $\varphi \in
C_0^{\infty}(\re^n)$ with $F \varphi (\xi) >0$ for $\xi \in
O(\xi_0)$ and $F \varphi (\xi_0)=1 $. Then it is easy to verify that
$A(D) \varphi(x)= \infty$. On the other hand, for $\varphi \in
\Psi_{G,p}(\re^n)$ the integral in Eq. (\ref{4}) is convergent due
to the compactness of $\mbox{supp } F\varphi  \subset G$.
We define the operator $A(-D)$ acting in the space
$\Psi_{-G,q}^{'}(\re^n)$ by the duality formula
\begin{equation}
\label{A(-D)} <A(-D)f,\varphi>\  =\ <f, A(D) \varphi>, \ f \in
\Psi_{-G,q}^{'}(\re^n),\ \varphi \in \Psi_{G,p}(\re^n).
\end{equation}
\begin{lemma}
The spaces $ \Psi_{G,p}(\re^n)$ and $\Psi_{-G,q}^{'}(\re^n)$ are
invariant with respect to the action of an arbitrary
pseudo-differential operator $A(D)$ whose symbol is continuous in
$G$. Moreover, if $A(\xi) \kappa_N(\xi)$ is a multiplier on $L_p$
for every $N\in \nat$, then this operator acts continuously.
\end{lemma}

\begin{remark}
In the case $p=2$ an arbitrary pseudo-differential operator whose
symbol is continuous in $G$ acts continuously without the additional
condition for $A(\xi) \kappa_N(\xi)$ to be multiplier in $L_2$ for
every $N\in\nat$.
\end{remark}
%%%%%%%
%%Subdiffusion
%

\subsection{Subdiffusion processes.}
\label{subdiffusion} As is known \cite{GLU,MK}, a (sub-)diffusion
process is governed by the fractional order partial differential
equation
\begin{equation}
\label{subdif} D_{\ast}^{\beta}u(t,x) = \mathcal{A}(D) u(t,x), t>0,
x \in \R^n,
\end{equation}
where $D_{\ast}^{\beta}$ is the Caputo fractional derivative of
order $\beta \in (0,1],$ and $\mathcal{A}(D), \, D=(D_1,...,D_n), \,
D_j= -i {\partial \over {\partial x_j}}, \, j=1,...,n,$ some
elliptic pseudo-differential operator.

%alteration{Text in this paragraph is revised}
Many diffusion processes driven by a Brownian motion can be
described by equation (\ref{subdif}) with a second order elliptic
differential operator $\mathcal{A}(D)$ and $\beta = 1.$ L\'evy
stochastic processes (which include jumps) also connected with
(\ref{subdif}) and an elliptic pseudo-differential operator
$\mathcal{A}(D)$ (see, e.g. \cite{Applebaum}). In particular, if
particle jumps are given by a symmetric L\'evy stable distribution
with infinite mean square displacement then $\mathcal{A}(D)$ is a
hyper-singular integral, defined as the inverse to the Riesz-Feller
fractional order ($0<\alpha < 2$) operator (for details see
\cite{SKM}). A wide variety of non-Gaussian stochastic
(subdiffusive) processes lead to equation (\ref{subdif}) with $0 <
\beta < 1$ (see \cite{MK,MK2004}). For diffusion governed by
distributed order differential equations see \cite{AUS,UmStan2006}.
The parameter $\beta$ determines the sub-diffusive mode, which is
slower than the classical free diffusion.

Generalizing this approach we will say that the $\{\beta_k, \mu,
\nu\}$-diffusion mode in the time interval $(T_{k-1},T_k)$ is
governed by the equation
\begin{equation}
\label{difmode} \mathcal{D}_{\ast \{\mu,\nu\}}^{\beta_k} u(t,x) =
\mathcal{A}(D) u(t,x), \, t\in (T_{k-1},T_k), \,  x \in \R^n.
\end{equation}
The entire process then can be described by the equation
\begin{equation}
\label{difmode1} \mathcal{D}_{\ast \{\mu,\nu\}}^{\beta(t)} u(t,x) =
\mathcal{A}(D) u(t,x), \, t>0, \,  x \in \R^n,
\end{equation}
where $\mathcal{D}_{\ast \{\mu,\nu\}}^{\beta(t)}$ is the variable
fractional order operator with the kernel
$K^{\beta(t)}_{\{\mu,\nu\}}(t, \tau)$ in   (\ref{KERNEL}).

In Section 4 we will prove the existence of solutions to the initial
value problem defined by the differential equation (\ref{difmode1}).
%%%%%%%%%%
%%  #3
%
\section{Changing of modes: 'short-range' and 'long-range' memories}

We call the triplet $\{\beta_k,\mu,\nu\}$ {\it admissible} if $0 <
\beta_k \le 1$ and $(\mu, \nu) \in \Pi$.  Diffusion in complex
heterogeneous media is accompanied by frequent changes of diffusion
modes. It is known that a particle undergoing non-Markovian movement
possesses a memory of past (see \cite{MK,Zaslavsky}). Protein
diffusion in cell membrane, as is recorded in
\cite{SaxtonJacobson97,Saxton01} is anomalous diffusion.
Descriptions of this process using random walks also shows the
presence of non-Markovian type memory \cite{AUS,GM,LSAT_05}. It
turns out, there is another type of memory noticed first by  Lorenzo
and Hartley in their paper \cite{LorenzoHartley} in some particular
cases of $\mu$ and $\nu$. This kind of memory arises when the
diffusion mode changes.

In this section we study a special case of this phenomenon where
there is a single change of diffusion mode, that is, a sub-diffusion
mode given by an admissible triplet $\{\beta_1,\mu,\nu\}$ changes to
a sub-diffusion mode corresponding to another admissible triplet
$\{\beta_2, \mu, \nu\}$ at some particular time $T$.
\begin{definition}
\label{DEF} Let $\{\beta_1, \mu, \nu\}$ and $\{\beta_2, \mu, \nu\}$
be two admissible triplets. Assume the diffusion mode is changed at
time $t=T$ from $\{\beta_1, \mu, \nu\}$-mode to $\{\beta_2, \mu,
\nu\}$-mode. Then the process is said to have a 'short-range' (or
short) memory, if there is a finite $T^{\ast} > T$ such that for all
%$t > T^{\ast}$ holds the mode $\{\beta_1, \mu, \nu\}$ and for
$t > T^{\ast}$ $\{\beta_2, \mu, \nu\}$-mode holds. Otherwise, the
process is said to have a 'long-range' (or long) memory.
\end{definition}
\begin{remark}
According to definition (\ref{DEF}), a diffusion mode has a long
memory if the influence of the old diffusion mode never vanishes,
even though the diffusion mode is changed, i.e. the particle does
not forget its past. In the case of  short memory, the particle
remembers the old mode for some critical time, and then forgets it
fully, recognizing the new mode.
\end{remark}
\begin{theorem}
\label{thmweak} Let $\nu > 0$ and $\mu \neq 0$. Assume the
$\{\beta_1, \mu, \nu\}$-diffusion mode is changed at time $t=T$ to
the $\{\beta_2, \mu, \nu\}$-diffusion mode. Let $T^{\ast}=T/\mu$ and
$t^{\ast}=T/(\mu + \nu).$ Then the process has a short memory.
Moreover,
\item{(i)}
$\{\beta_1, \mu, \nu\}$-diffusion mode holds for all $0<t<t^{\ast};$
\item{(ii)}
$\{\beta_2, \mu, \nu\}$-diffusion mode holds for all $t>T^{\ast};$
\item{(iii)}
a mix of both $\{\beta_1, \mu, \nu\}$ and $\{\beta_2, \mu,
\nu\}$-diffusion modes holds for all $t^{\ast}<t<T^{\ast}.$
\end{theorem}

{\it Proof.} Let $\beta (s)=\beta_1$ for $0<s<T$ and $\beta
(s)=\beta_2$ for $s>T.$ Assume $\nu >0$. Denote $s=\mu t+\nu \tau.$
So, the $\{\beta_1, \mu, \nu\}$-diffusion mode holds if $\mu t+\nu
\tau <T$. Let $0<t<t^{\ast}= T/(\mu+\nu)$. Then for every $\tau \in
(0,t)$ we have $\mu t + \nu \tau < (\mu + \nu)t < T$. This means
that the order operator $\beta(s)$  in $\mathcal{D}_{\ast
\{\mu,\nu\}}^{\beta(t)}$  takes the value $\beta_1$ giving (i). If
$t>T/\mu$ then for all $\tau>0$, $\mu t+\nu\tau >T$. Hence,
$\beta(s)=\beta_2,$ obtaining (ii). Now assume $T/(\mu +
\nu)<t<T/\mu.$ Denote $\tau_0=(T-\mu t)/\nu$. Obviously $\tau_0 >
0.$ It follows from $(\mu+\nu)t>T$ dividing by $\nu$ that $t>T/\nu-
t\mu/\nu=\tau_0$, i.e. $0<\tau_0 < t.$ It is easy to check that if
$0<\tau<\tau_0$ then $\mu t +\nu \tau \in (\mu t, T) \subset (0,T),$
giving $\beta(s)=\beta_1,$ while if $\tau_0 <\tau < t$ then $\mu t
+\nu \tau \in (T, (\mu+\nu) t) \subset (T,\infty),$ giving
$\beta(s)=\beta_2.$ Hence,  in this case the mix of both $\{\beta_1,
\mu, \nu\}$ and $\{\beta_2, \mu, \nu\}$-diffusion modes is present.
\eproof

\begin{theorem}
\label{thmstrong} Let $\nu < 0$ and $\mu + \nu \neq 0$. Assume the
$\{\beta_1, \mu, \nu\}$-diffusion mode is changed at time $t=T$ to
the $\{\beta_2, \mu, \nu\}$-diffusion mode. Let
${t^{\ast}}^{'}=T/\mu$ and ${T^{\ast}}^{'}=T/(\mu + \nu).$ Then the
process has a short memory. Moreover,
\item{$(i^{'})$}
$\{\beta_1, \mu, \nu\}$-diffusion mode for all $0<t<{t^{\ast}}^{'};$
\item{$(ii^{'})$}
$\{\beta_2, \mu, \nu\}$-diffusion mode holds for all
$t>{T^{\ast}}^{'};$
\item{$(iii^{'})$}
a mix of both $\{\beta_1, \mu, \nu\}$ and $\{\beta_2, \mu,
\nu\}$-diffusion modes holds for all
${t^{\ast}}^{'}<t<{T^{\ast}}^{'}.$
\end{theorem}

{\it Proof.} Let $\nu<0$. Assume again $\beta (s)=\beta_1$ for
$0<s<T$ and $\beta (s)=\beta_2$ for $s>T.$ As in the previous
theorem, denote $s=\mu t+\nu\tau.$ First we notice that if
$0<t<T/\mu$ then $\mu t + \nu \tau <T$, which implies
$\beta(s)=\beta_1$, giving ($i^{'}$). Now let $t > T/(\mu+\nu)$ be
any number. Then for $0 < \tau < t$ we have $\mu t + \nu \tau > T$,
which yields $\beta(s)=\beta_2.$ So, we get ($ii^{'}$). Now assume
$T/\mu<t<T/(\mu+\nu).$ Again denote $\tau_0=(T-\mu t)/\nu$.
Obviously $\tau_0 > 0.$ It follows from $(\mu+\nu)t < T$ dividing by
$\nu < 0$ that $t>T/\nu- t\mu/\nu=\tau_0$, i.e. $0<\tau_0 < t.$ It
is easy to check that if $0<\tau<\tau_0$ then $\mu t +\nu \tau \in
(T,\mu t) \subset (T, \infty),$ giving $\beta(s)=\beta_2,$ while if
$\tau_0 <\tau < t$ then $\mu t +\nu \tau \in ((\mu+\nu) t, T)
\subset (0,T),$ giving $\beta(s)=\beta_1.$ Hence,  in this case the
mix of both $\{\beta_1, \mu, \nu\}$ and $\{\beta_2, \mu,
\nu\}$-diffusion modes is present, obtaining ($iii^{'}$). \eproof

\begin{corollary}
\label{thmnuzero} Let $\nu = 0$ and $\mu \neq 0$. Assume the
$\{\beta_1, \mu, \nu\}$-diffusion mode is changed at time $t=T$ to
the $\{\beta_2, \mu, \nu\}$-diffusion mode. Let $T^{\ast}=T/\mu$.
Then the process has a short memory. Moreover,
\item{(a)}
for all $0<t<T^{\ast}$ there holds $\{\beta_1, \mu, \nu\}$-diffusion
mode;
\item{(b)}
for all $t>T^{\ast}$ there holds $\{\beta_2, \mu, \nu\}$-diffusion
mode.
\end{corollary}

{\it Proof.} If $\nu=0$ then we have $\beta(s)=\beta(\mu t) =
\beta_1$ for $t < T/\mu$ and $\beta(s)=\beta_2$ for $t>T/\mu$.
\eproof

\begin{corollary}
\label{long} Let $\mu = 0$ or $\mu+\nu =0$.  Assume the $\{\beta_1,
\mu, \nu\}$-diffusion mode is changed at time $t=T$ to the
$\{\beta_2, \mu, \nu\}$-diffusion mode. Then the process has the
long memory.
\end{corollary}

{\it Proof.} According to the structure of LH-parallelogram $\mu =0$
implies $\nu > 0.$ In this case $T^{\ast}=\infty$. If $\mu + \nu =
0$ then $\nu <0$ and $t^{\ast}=\infty.$ In both cases we a have long
memory effect. \eproof

\begin{remark}
Notice, that if $\nu=0$ then there is no intervals of mix of modes.
Moreover, if $\nu=0, \, \mu=1,$ then $T^{\ast}=t^{\ast}=T.$ In this
sense we say that a process has no memory. For all points  $\{(\mu,
\nu)\}$ except $\{\mu=0,0 \le \nu \le 1\}$ and $\{\nu<0,
\mu+\nu=0\},$ the operator $ D_{\ast \{\mu,\nu\}}^{\beta(t)}$ has a
short memory. The memory is stronger in the region $\nu <0$ and
weaker in $\nu
>0$. On the line $\mu + \nu = 1$ we have $t^{\ast}=T <
T^{\ast}.$ The lines $\mu=0, \, \nu \geq 0$ and $\mu + \nu =0$
identify the long range memory.
\end{remark}

%
%  #4
%
\section{The Cauchy problem for variable order differential equations}
In this section we study the Cauchy problem for variable order
differential equations with a piecewise constant order function
$\beta(t)=\sum_{k=0}^{N} \mathcal{I}_k \beta_k$, where
$\mathcal{I}_k$ is the indicator function of $[T_{k},T_{k+1}).$ We
assume that the diffusion mode change times $T_1, T_2,...,T_{N}$ are
known, and set $T_0=0,\, T_{N+1}=\infty.$ We assume that the
solution of the initial value problem for the VOPDE (\ref{difmode1})
is
% alteration {footnote added}
continuous\footnote{in the topology of $\Psi_{G,p}(\R^n)$ (or
$\Psi_{-G,p}^{'}$).} when the diffusion mode changes.

Thus, the Cauchy problem is formulated in the form
\begin{equation}
\label{GenEq} \mathcal{D}_{\ast \{\mu,\nu\}}^{\beta(t)} u(t,x) =
\mathcal{A}(D) u(t,x), \, t>0, \,  x \in \R^n,
\end{equation}
\begin{equation}
\label{initial} u(0,x)= \vf (x),
\end{equation}
\begin{equation}
\label{continuity} u(T_k^{\ast}-0,x) = u(T_{k}^{\ast}+0, x), \,
k=1,...,N, \, x \in \R^n,
\end{equation}
where $\mathcal{A}(D)$ is a pseudo-differential operator with a
continuous symbol $A(\xi), \, \xi \in R^n,$ and $T^{\ast}_k$ are
actual mode change times. It follows from Theorems \ref{thmweak} and
\ref{thmstrong} that $T^{\ast}_{k}$ are defined through
$T_k/(\mu+\nu)$ and $T_k/\mu.$

We note that,  since the integration operator order $\beta$ depends
on the variable $t$,  a variable order analog of the integration
operator becomes
\begin{equation}
\label{frintegrationvar} J^{\beta(t)}_{\{\mu, \nu\}} f(t)=
\int_0^{t} \frac{ (t-\tau)^{\beta (\mu t +\nu \tau) - 1}
 f(\tau)}{{\Gamma (\beta (\mu t +\nu \tau))}} d \tau,
%\]
\end{equation}
which we call {\em a variable order integration operator}.

%alteration{This paragraph is added by the request of referee}
For further purpose we recall the definition of the Mittag-Leffler
function \cite{Djr93,Pollard48} in the power series form
\[
E_{\beta}(z)=\sum_{n=0}^{\infty}\frac{z^n}{\Gamma(\beta n+1)}, \,
z\in C^1.
\]
Obviously, $E_{\beta}(z)=e^z,$ if $\beta =1.$ For all $\beta
> 0$ $E_{\beta}(z)$ is an entire function of type 1 and order $1/\beta.$
Note that $E_{\beta}(-t), ~ t>0,$ is completely monotone
\cite{Pollard48}, and has asymptotics $E_{\beta}(-t) = O(t^{-1}), ~
t \rightarrow \infty.$
\begin{lemma}
Assume $0<\beta_{\ast} = min_{0 \leq j \leq N} \{\beta_j\}$ and $[k
\beta_{\ast}]$ is the integer part of $k \beta_{\ast}.$ Let $v(t)$
be a function continuous in $[0,\infty)$. Then for arbitrary $T > 0$
and every $k = 1,2,... $ the estimate
\begin{equation}
\label{estimate} max_{0 \leq t \leq T} |J^{\beta(t)k} v(t)| \leq
\frac{{[\psi}(T)]^{k}}{[k \beta_{\ast}+1]!} max_{0\leq t \leq
T}|v(t)|
\end{equation}
holds with
\[
\psi(\tau)=
  \left\{ \begin{array}{ll}
    \tau^{\beta_{\ast}}, & 0<\tau<1, \\
    {\ds
    \tau}, & \tau \geq 1. \\
    \end{array} \right.
\]
\end{lemma}

{\it Proof.} Let $v(t)$ be a function continuous in $[0,\infty)$.
For $k$ large enough, so that $\beta_{\ast} k \geq 2$ we have $min
\Gamma (k \beta(\mu t + \nu \tau)) = \Gamma (k \beta_{\ast})$.
Taking this into account, for all such $k$ and for all $t \in (0,T]$
we obtain the estimate
\[
|J^{\beta(t)k} v(t)| = |\int_0^t\frac{(t-\tau)^{k \beta(\mu t + \nu
\tau)-1} v(\tau) d \tau}{\Gamma(k \beta(\mu t + \nu \tau))}| \leq
\frac{[\psi(T)]^{k}}{\Gamma(k \beta_{\ast}+1)} max_{0\leq t \leq
T}|v(t)|,
\]
and hence, the estimate in Eq.  (\ref{estimate}). \eproof

Let $t_{cr,j}=T_j/(\mu+\nu), j=1,...,N,$ be critical points
corresponding to the mode change times $T_j, j=1,...,N.$ We accept
the conventions $t_{cr,0}=0, t_{cr,N+1}=\infty.$ Let $E_{\beta}(z)$
be the Mittag-Leffler function with parameter $\beta \in(0,1]$. Now
we introduce the symbols which play an important role in the
representation of a solution. Let
\begin{equation}
\label{sj} {S}_j(t,\xi)=E_{\beta_j}((t-t_{cr,j})^{\beta_j}A(\xi)),
\, t\geq t_{cr,j}, \, j=0,...,N,
\end{equation}
and
\begin{equation}
\label{mk} M_k(t,\xi) = S_k(t-t_{cr,k}, \xi) \prod_{j=0}^{k-1} S_j
(t_{cr,j+1}-t_{cr,j}, \xi), \, t \geq t_{cr,k}, \, k=1,...,N.
\end{equation}
Further, we define recurrently the symbols
\[
\mathcal{R}_1(t, \xi) = - \frac{1}{\Gamma(1-\beta_1)}
\int_0^{t_{cr,1}} \frac{\frac{\partial}{\partial
\tau}S_0(\tau,\xi)}{(t-\tau)^{\beta_1}}d \tau = \frac{-\beta_0
A(\xi)}{\Gamma(1-\beta_{1})} \int_{0}^{t_{cr,1}}
\frac{E_{\beta_0}^{'}(\tau^{\beta_0} A(\xi))d
\tau}{\tau^{1-\beta_0}(t-\tau)^{\beta_{1}}},
\]
\[
P_{-1}(t,\xi) \equiv 0, \, P_0(t,\xi) \equiv 1, \, P_1(t,
\xi)=\int_{t_{cr_1}}^{t} S_j(t+t_{cr,1}-\tau, \xi) \, _{t_{cr,1}}
D_{\tau}^{1-\beta_1} \mathcal{R}_1(\tau, \xi)d \tau,
\]
and if
\[
P_j(t, \xi)=\int_{t_{cr_j}}^{t} S_j(t + t_{cr,j}-\tau, \xi) \,
_{t_{cr,j}} D_{\tau}^{1-\beta_j} \mathcal{R}_j(\tau, \xi)d \tau,
\]
is defined for $t \geq t_{cr,j}$ and for all $j \leq k-1,$ then for
$t \geq t_{cr,k},$

\vspace{.2cm}

$ \mathcal{R}_k (t,\xi) = $
\begin{equation}
\label{rk}  - \frac{1}{\Gamma(1-\beta_k)}\sum_{j=0}^{k-1}
\int_{t_{cr,j}}^{t_{cr,j+1}} \frac{\frac{\partial}{\partial \tau}
[M_j(\tau,\xi)+S_j(\tau-t_{cr,j},\xi)
P_{j-1}(t_{cr,j},\xi)+P_j(\tau,\xi)]}{(t-\tau)^{\beta_k}}d \tau,
\end{equation}
for $k=2,...,N.$

\subsubsection{The case $\nu=0$.}

\begin{theorem}
\label{Solutiontheorem1} Assume $\nu = 0$ and $\varphi \in
\Psi_{G,p}(\R^n).$ Then the Cauchy problem
(\ref{GenEq})-(\ref{continuity}) has a unique solution $u(t,x) \in
C([0,T],\Psi_{G,p}(\R^n))$, $T < \infty,$ which is represented in
the form $u(t,x)= \mathcal{S}(t,D)\varphi (x),$ where $\mathcal{S}
(t,D)$ is the pseudo-differential operator with the symbol
\[
\mathcal{S}(t,\xi)= \mathcal{I}^{'}_0 S_0(t,\xi)  + \sum_{k=1}^N
\mathcal{I}^{'}_k (t) \, \, \{ \, M_k(t,\xi) \, \, \, + \]
\[S_k(t,\xi) \int_{t_{cr,k-1}}^{t_{cr,k}} S_{k-1} (t_{cr,k} +
t_{cr,k-1}- \tau, \xi) \, \, \, \, _{t_{cr,k-1}}D_{\tau}^{1-\beta_k}
\mathcal{R}_k (\tau,\xi) d \tau \, \, \, +
\]
\begin{equation}
\label{solrepresentation} \int_{t_{cr,k}}^t S_k (t + t_{cr,k}- \tau,
\xi) \, \, \, \, _{t_{cr,k}}D_{\tau}^{1-\beta_k} \mathcal{R}_k
(\tau,\xi) d \tau \} \, ,
\end{equation}
where $\mathcal{I}^{'}_k=\mathcal{I}_{[t_{cr,k},t_{cr,k+1})}(t), \,
k=0,...,N,$ are indicator functions of the intervals
$[t_{cr,k},t_{cr,k+1}),$ $k=0,...,N$; $S_j(t,\xi), \, j=0,...,N,$
$M_k(t,\xi)$ and $\mathcal{R}_k(t,\xi), \, k=1,...,N,$ are defined
in (\ref{sj}), (\ref{mk}) and (\ref{rk}), respectively.
\end{theorem}

{\it Proof.} It is not hard to verify that
\begin{equation}
\label{fm} J^{\beta(t)}_{\{\mu,0\}} \mathcal{D}_{\ast
\{\mu,0\}}^{\beta(t)} u(t,x) = \sum_{k=0}^{N} \mathcal{I}^{'}_k
J^{\beta_k}D_{\ast}^{\beta_k} u(t,x) = u(t,x) - \sum_{k=0}^{N}
\mathcal{I}^{'}_k u(t_{cr,k},x) + g (t,x),
\end{equation}
where
\[
g(t,x) = \sum_{k=1}^{N} \mathcal{I}^{'}_k \frac{t^{\beta_k}}{\Gamma
(1+\beta_k)} \sum_{j=0}^{k-1} \,_{t_{cr,j}} \mathcal{D}_{\ast
\{\mu,0\}}^{\beta_k} u(t_{cr, j+1},x).
\]
Multiplying both sides of equation (\ref{GenEq}) by
$J^{\beta(t)}_{\{\mu,0\}}$ and applying the formula (\ref{fm}), we
obtain
\begin{equation}
\label{geneq} u(t,x) - \sum_{k=0}^{N} \mathcal{I}^{'}_k J^{\beta_k}
\mathcal{A}(D) u(t,x) =  \sum_{k=0}^{N} \mathcal{I}^{'}_k
u(t_{cr,k},x) - g(t,x).
\end{equation}
Let $t \in (0,t_{cr,1}).$ Then $\beta(\mu t)=\beta_0$ and $g(t,x)
\equiv 0.$ In this case taking into account the initial condition
(\ref{initial}),
 we can rewrite equation (\ref{geneq}) in the form
\begin{equation}
u(t,x)- J^{\beta_0} \mathcal{A}(D) u(t,x) = \vf(x), \, 0<t<t_{cr,1}.
\end{equation}
The obtained equation can be solved by using the iteration method.
Determine the sequence of functions $\{u_0(t,x),...,u_m(t,x)\}$ in
the following way. Let $u_0(t,x)= \vf (x)$ and by iteration
\begin{equation}
\label{iteration} u_m(t,x) = J^{\beta_0} \mathcal{A}(D)u_{m-1}(t,x)
+ \vf(x), \, m=1,2,...
\end{equation}
We show that this sequence is convergent in the topology of the
space $C[0,T;\Psi(R^n)]$ and its limit is a solution to the Cauchy
problem   (\ref{GenEq}),(\ref{initial}). Moreover, this solution can
be represented in the form of functional series
\begin{equation}
\label{funcseries} u(t,x)= \sum_{k=0}^{\infty} J^{\beta_0 k}
\mathcal{A}^k(D)\vf(x).
\end{equation}
Indeed, it follows from the iteration process  (\ref{iteration})
that
\begin{equation}
\label{u_m} u_m (t,x)= J^{\beta(t)m} \mathcal{A}^m(D) \vf (x) +
J^{\beta(t)(m-1)} \mathcal{A}^{m-1}(D) \vf (x) +...+ \vf (x).
\end{equation}
Now we estimate $u_m(t,x)$ applying Lemma \ref{estimate} term by
term in the right hand side of  (\ref{u_m}). Indeed, let $N \in
\nat.$ Then taking into account the fact that the Fourier transform
in $x$ commutes with $J^{\beta(t)}$, we have
\[
max_{[0,T]}p_N(J^{\beta_0 k} \mathcal{A}^k(D)\vf(x)) \leq
\frac{\psi^{k-1}(T)}{[k \beta_0]!} p_N (\mathcal{A}\vf(x)).
\]
Further, since $A(\xi)$ is continuous on $G$ there exists a constant
$C_N > 0$, such that $ max_{\xi \in supp \, \kappa_N} |A(\xi)| \leq
C_N, $ or, by induction $ max_{\xi \in supp \, \kappa_N} |A^k(\xi)|
\leq C^k_N. $ Hence, for every $N \in \nat$ we have
\begin{equation}
\label{est} p_N(J^{\beta(t)k} \mathcal{A}^k(D)\vf(x)) \leq \|\vf\|_p
\frac{C_N^{k-1} \, \psi^{k-1}(T)}{[k \beta_0]!}.
\end{equation}
It follows from (\ref{est}) that
\[
max_{[0,T]}p_N (u_m(t,x)) \leq \|\vf(x)\|_p \sum_{k=0}^m \frac{C_N^k
\psi^k(T)}{\Gamma(\beta_0 k +1)}
\]
\[
\leq C \|\vf(x)\|_p E_{\beta_0}(C_N \, \psi(T)), \, N=1,2,...,
\]
where $E_{\beta_0}(\tau)$ is the Mittag-Leffler function
corresponding to $\beta_0$. As far as the right hand side of the
latter does not depend on $m$, we conclude that $u_m(t,x)$ defined
in (\ref{u_m}) is convergent. Again making use of Lemma
\ref{estimate} we have
\begin{equation}
\label{conver} p_N (u(t,x)-u_m(t,x)) \leq \|\vf(x)\|_p
\sum_{k=m+1}^{\infty} \frac{C_N \psi^k(T)}{\Gamma(\beta_0 k +1)}, \,
N=1,2,....
\end{equation}
The function
\[
\mathcal{R}_m(\eta) = \sum_{k=m+1}^{\infty}
\frac{\eta^{k}}{\Gamma(\beta_0 k +1)}
\]
on the right side of equation (\ref{conver}) is the residue in the
power series representation of the Mittag-Leffler function
$E_{\beta_0}(\eta),$ and, hence,
%$E_{\beta_0}(\eta)$ is an entire function, i.e.
$\mathcal{R}_m(\eta) \rightarrow 0$, when $m \rightarrow \infty$ for
any real (or even complex) $\eta.$ Consequently, $u_m(t,x)
\rightarrow u(t,x)$ for every $N=1,2,...,$, that is in the inductive
topology of the space $C([0, \infty), \Psi_{G,p})$. Thus, $u(t,x)
\in C([0, \infty), \Psi_{G,p})$ is a solution. Moreover, it is
readily seen that $u(t,x)$ in  (\ref{funcseries}) can be represented
through the pseudo-differential operator $S(t,D)$ with the symbol
$S_0(t,\xi)= E_{\beta_0} (t^{\beta_0} \mathcal{A}(\xi))$ in the form
$u(t,x)=u_0(t,x)=S_0(t,D)\vf(x),\, t\in (0, t_{cr,1})$. By
construction $u(t,x)$ is unique and continuous in $t$. So, $\lim_{t
\rightarrow t_{cr,1}-0}u(t,x)= E_{\beta_0}(t^{\beta_0}_{cr,1}
\mathcal{A}(D))\vf(x) $ exists in $\Psi_{G,p}(\R^n).$ Further we
extend $u(t,x)$ to $[t_{cr,1},t_{cr,2}).$ Equation (\ref{GenEq}) in
the interval $(t_{cr,1},t_{cr,2})$ reads
\[
D_{\ast}^{\beta_1} u(t,x)=\mathcal{A}(D)u(t,x), \, t \in
(t_{cr,1},t_{cr,2}).
\]
Splitting the integration interval $(0,t)$ on the left hand side of
the last equation into subintervals $(0,t_{cr,1})$ and
$(t_{cr,1},t),$ we can rewrite it in the form
\begin{equation}
\label{difeq-1} _{t_{cr,1}}D_{\ast}^{\beta_1} u(t,x)=
\mathcal{A}(D)u(t,x) + F_1(t, x), \, t \in (t_{cr,1},t_{cr,2}),
\end{equation}
where
\[
F_1(t,x) = -\frac{1}{\Gamma(1-\beta_1)}\int_0^{t_{cr,1}}
\frac{\frac{\partial}{\partial \tau} u_0
(\tau,x)}{(t-\tau)^{\beta_1}} d\tau,
\]
Taking into account the fact $u_0(t,x)=S_0(t,D)\vf(x),$ it is not
hard to see that $F_1(t,x) = \mathcal{R}_1 (t, D)\vf(x),$ where
\[
\mathcal{R}_1 (t, \xi) = \frac{-\beta_0 A(\xi)}{\Gamma(1-\beta_{1})}
\int_{0}^{t_{cr,1}} \frac{E_{\beta_0}^{'}(\tau^{\beta_0}
A(\xi))}{\tau^{1-\beta_0}(t-\tau)^{\beta_{1}}} d\tau.
\]
Due to the continuity condition (\ref{continuity}), we have also
\begin{equation}
\label{cauchy-1} u(t_{cr,1},x)=u_0 (t_{cr,1},x)=
E_{\beta_0}(t^{\beta_0}_{cr,1}
\mathcal{A}(D))\vf(x)=S_0(t_{cr,1},D)\vf(x).
\end{equation}
In the general case, assuming that solution are found in the
intervals $[0,t_{cr,1}), ... , $ $ [t_{cr,k-1},t_{cr,k}),$ we have
the following inhomogeneous Cauchy problem for the interval
$(t_{cr,k},t_{cr,k+1})$
\begin{equation}
\label{difeq-k} _{t_{cr,k}}D_{\ast}^{\beta_k} u(t,x)=
\mathcal{A}(D)u(t,x) + F_k(t,x), \, t \in (t_{cr,k},t_{cr,k+1}),
\end{equation}
\begin{equation}
\label{cauchy-k} u(t_{cr,k},x)=u_{k-1} (t_{cr,k},x),
\end{equation}
where
\[
F_k(t, x) = - \frac{1}{\Gamma(1-\beta_k)} \sum_{j=0}^{k-1} \,
\int_{t_{cr,j}}^{t_{cr,j+1}} \frac{\frac{\partial}{\partial \tau}
u_j(\tau,x)}{(t-\tau)^{\beta_k}} d \tau,
\]
It is not hard to verify that $F_k(t,x)$ can be represented in the
form $\mathcal{R}_k(t,D) \vf (x)$ with a pseudo-differential
operator $\mathcal{R}_k(t,D)$ whose symbol is given in (\ref{rk}).
%a combination of operators $\,_{t_{cr,j}}D_{\ast}^{\beta_m}
%E_{\beta_l} ((t_{cr,j+1}-t_{cr,j})^{\beta_l}A(D)).$
A unique solution to (\ref{difeq-k}),(\ref{cauchy-k}) can be found
applying the fractional Duhamel principle (see \cite{UmSay2006})
\[
u_k(t,x)= S_k(t,D)
%E_{\beta_k}((t-t_{cr,k})^{\beta_k} \mathcal{A}(D))
u_{k-1}(t_{cr,k},x) + \]
\[
\int_{t_{cr,k}}^t S_k(t- (\tau-t_{cr,k}), D )
_{t_{cr,k}}D_{\tau}^{1-\beta_k} F_k(\tau, x)d \tau,
\]
\[
k=1,...,N, \, t_{cr,k} < t < t_{cr,k+1}.
\]
Now taking into account that
\[
u_{k-1}(t_{cr,k},x) =  \left[\prod_{j=0}^{k-1} S_j
(t_{cr,j+1}-t_{cr,j}, D)\right] \vf(x) +
\]
\[
\int_{t_{cr,k-1}}^{t_{cr,k}} S_{k-1}(t_{cr,k}- (\tau-t_{cr,k-1}), D
) _{t_{cr,k-1}}D_{\tau}^{1-\beta_{k-1}} F_{k-1}(\tau, x)d \tau
\]
we arrive at  (\ref{solrepresentation}). \eproof

\begin{remark}
Assume in equation (\ref{GenEq}) $\beta(t)=\beta$, where $\beta$ is
a constant in $(0,1]$. Then the representation formula
(\ref{solrepresentation}) is reduced to
$$
u(t,x)=E_{\beta}(t^{\beta} \mathcal{A(D)}) \vf(x),
$$
which coincides with the result of \cite{GLU}.
\end{remark}

%Dual Theorem
Applying the technique used in the paper \cite{GLU} and the duality
of the spaces $\Psi_{G,p}(\R^n)$ and $\Psi_{-G,q}^{'}(\R^n)$ we can
prove the following theorem.

\begin{theorem}
\label{dual} Assume $\nu = 0$ and $\varphi \in
\Psi^{'}_{-G,q}(\R^n).$ Then the Cauchy problem
(\ref{GenEq})-(\ref{continuity}) (with '-D' instead of 'D') has a
unique weak solution $u(t,x) \in C([0,T],\Psi^{'}_{-G,q}(\R^n))$, $T
< \infty,$ which is represented in the form
\[
u(t,x)= \mathcal{I}_0^{'} S_0(t,-D)\vf(x) +  \sum_{k=1}^N
\mathcal{I}^{'}_k (t) \, \, \{ \, M_k(t, -D) \vf (x) +
\]
\[
S_k(t,-D) \int_{t_{cr,k-1}}^{t_{cr,k}} S_{k-1} (t_{cr,k} +
t_{cr,k-1}- \tau, -D) \, \, \, \, _{t_{cr,k-1}}D_{\tau}^{1-\beta_k}
\mathcal{R}_k (\tau,-D) \vf(x) d \tau \,  \, \, +
\]
\begin{equation}
\label{dualrepresentation} \int_{t_{cr,k}}^t S_k (t + t_{cr,k}-
\tau, -D) \, \, \, \, _{t_{cr,k}}D_{\tau}^{1-\beta_k} \mathcal{R}_k
(\tau, -D) \vf (x)d \tau \, \},
\end{equation}
\end{theorem}

\begin{corollary}
\label{cor} If $\nu = 0$ then the fundamental solution of equation
(\ref{GenEq}) with the continuity conditions in  (\ref{continuity})
is represented in the form
\[
U(t,x)= \mathcal{I}^{'}_0 (t) \frac{1}{(2\pi)^n} \int_{R^n}
E_{\beta_0}(t^{\beta_0}\mathcal{A}(-{\xi}))d\xi +
\]
\[
\sum_{k=1}^N \mathcal{I}^{'}_k (t) \frac{1}{(2\pi)^n}\int_{R^n} \,
\, \{ \, E_{\beta_k}((t-t_{k})^{\beta_k} \mathcal{A}(-\xi))
 \prod_{j=0}^{k-1} E_{\beta_j} ((t_{j+1}-t_{j})^{\beta_j} \mathcal{A}(-\xi)) +
\]
\[
E_{\beta_k} ((t- t_{k})^{\beta_k} A(-\xi)) \int_{t_{k-1}}^{t_{k}}
E_{\beta_{k-1}} ((t_{k} - \tau)^{\beta_{k-1}} A(-\xi)) \, \, \, \,
_{t_{k-1}}D_{\tau}^{1-\beta_{k-1}} \mathcal{R}_{k-1} (\tau, -\xi) d
\tau +
\]
\begin{equation}
\label{fundsolution}
%\sum_{k=0}^{N} \mathcal{I}^{'}_k (t) \frac{1}{(2\pi)^n}\int_{R^n}
\int_{t_{k}}^t E_{\beta_k} ((t- \tau)^{\beta_k} A(-\xi)) \, \, \, \,
_{t_{k}}D_{\tau}^{1-\beta_k} \mathcal{R}_k (\tau, -\xi)
 d \tau \, \} e^{i x \xi}\, d \xi,
\end{equation}
where $t_j = t_{cr,j}.$ Moreover, $U(t,x) \in \Psi_{-G,q}^{'}(\R^n)$
for every fixed $t>0$.
\end{corollary}
%%%

\subsection{The case $-1< \nu \leq 1$}

The solution $u(t,x)=\mathcal{S}(t,D)\vf(x)$ obtained in Theorem
\ref{Solutiontheorem1} in the case $\nu=0$ has the structure
$u(t,x)=\Psi_1(t,D)\vf(x) + \Psi_2(t,D)\vf(x),$ where $\Psi_1 (t,D)$
and $\Psi_2(t,D)$ are operators with symbols
\[
\Psi_1(t,\xi)= \mathcal{I}^{'}_0 S_0(t,\xi)  + \sum_{k=1}^N
\mathcal{I}^{'}_k (t) \, M_k(t,\xi),
\]
and

$ \Psi_2(t,\xi)= $
\[
\sum_{k=1}^N \mathcal{I}^{'}_k(t) \{ S_k(t,\xi)
\int_{t_{cr,k-1}}^{t_{cr,k}} S_{k-1} (t_{cr,k} + t_{cr,k-1}- \tau,
\xi) \, \, \, \, _{t_{cr,k-1}}D_{\tau}^{1-\beta_k} \mathcal{R}_k
(\tau,\xi) d \tau \, \, \, +
\]
\begin{equation}
\label{psi2} \int_{t_{cr,k}}^t S_k (t + t_{cr,k}- \tau, \xi) \, \,
\, \, _{t_{cr,k}}D_{\tau}^{1-\beta_k} \mathcal{R}_k (\tau,\xi) d
\tau \}
\end{equation}
The term $v(t,x)=\Psi(t,D)\vf(x)$ reflects an effect of diffusion
modes, while the term $w(t,x)=\Psi_2(t,D)\vf(x)$ is connected with a
memory of the past. We note that this structure remains correct in
the general case $\nu \in (0,1]$ also, however, the symbols of
solution operators are further restructured depending on the
intervals of mixture of (two or more) modes. The theorems formulated
below concern some intervals free of mixed modes.
\begin{theorem}
\label{nupositive} Assume $\mu \neq 0, \, \mu + \nu \neq 0$ and
$\varphi \in \Psi_{G,p}(\R^n).$  Then there exists a number
$T^{\ast}>0$ and pseudo-differential operators
$\mathcal{P}^{\ast}(D)$ and $\mathcal{R}^{\ast}(t,D)$ with
continuous symbols, such that for $t > T^{\ast}$ the solution of the
Cauchy problem (\ref{GenEq})-(\ref{continuity}) coincides with the
solution of the Cauchy problem
\begin{equation}
\label{betaN} \,_{T^{\ast}}D_{\ast}^{\beta_N} u(t,x)= \mathcal{A}(D)
u(t,x) + f^{\ast}(t,x), \, t>T^{\ast}, \, x \in \R^n,
\end{equation}
\begin{equation}
\label{innupos} u(T^{\ast},x)= \vf^{\ast}(x), \, x \in \R^n.
\end{equation}
where $f^{\ast}(t,x) = \mathcal{R}^{\ast}(t,D) \vf(x)$ and
$\vf^{\ast}(x) = \mathcal{P}^{\ast}(D) \vf(x).$
\end{theorem}

{\it Proof.} Assume $\nu > 0$. Then as it follows from Theorem
\ref{thmweak} that the actual mode changes occur at times
$T^{\ast}_{j} = T_j/\mu$ and $t^{\ast}_{j} = T_j/(\mu + \nu), \,
j=1,...,N,$ if diffusion modes change at times $T_j, j=1,..,N.$
Obviously, $t^{\ast}_{1} < ... < t^{\ast}_{N}$ and $T^{\ast}_{1} <
... < T^{\ast}_{N}$ if $T_1<...<T_N.$ The order function $\beta(\mu
t +\nu \tau)$ under the integral in $D_{\ast \{\mu,
\nu\}}^{\beta(t)}$ takes the value $\beta_N$ for all $t >
T^{\ast}_{N}$ and $\tau > 0.$  Hence, the variable order operator on
the left side of (\ref{GenEq}) becomes $D_{\ast}^{\beta_N}$ if $t
> T^{\ast}_{N}$. Analogously it follows from Theorem \ref{thmstrong}
that  if $\nu < 0$ then $\beta(\mu t +\nu \tau)$ takes the value
$\beta_N$ for all $t > t^{\ast}_{N}$ and $\tau > 0.$ Thus, if $\nu
\neq 0$ then for all $t > T^{\ast} = max \{T^{\ast}_{N},
t^{\ast}_{N} \}$ and $0 < \tau < t $ we have $\beta(\mu t +\nu \tau)
= \beta_N.$ Similar to the case $\nu=0$, splitting the interval
$(0,t), \, \, t > T^{\ast},$ into subintervals, we can represent the
equation (\ref{GenEq}) in the form (\ref{betaN}). Further, from the
continuity condition (\ref{continuity}) we have $u(T^{\ast},x) =
\lim_{t \rightarrow T^{\ast}-0}v(t,x),$ where $v(t,x)$ is a solution
to the Cauchy problem for fractional order pseudo-differential
equations in sub-intervals of the interval $[0, T^{\ast}_{N})$
constructed by continuation. Therefore there exists an operator
$S^{\ast}(t, D),$ such that $v(t,x)=S^{\ast}(t,D)\vf(x).$ Denote
$\mathcal{P}^{\ast}(D)=S^{\ast}(T^{\ast},D).$ Then
$u(T^{\ast},x)=\mathcal{P}^{\ast}(D)\vf(x).$ This means that for
$t>T^{\ast}$ solutions of problems (\ref{GenEq})-(\ref{continuity})
and (\ref{betaN}),(\ref{innupos}) coincide. If $\nu =0$, then the
statement follows from Theorem \ref{Solutiontheorem1}. \eproof

\begin{theorem}
\label{nunegative} Assume $\varphi \in \Psi_{G,p}(\R^n).$ Then there
exists a number $t^{\ast}>0,$
%and a pseudo-differential operator $p^{\ast}(D)$,
such that for $0< t < t^{\ast}$ the solution of the Cauchy problem
(\ref{GenEq})-(\ref{continuity}) coincides with the solution of the
Cauchy problem
\begin{equation}
\label{beta0} D_{\ast}^{\beta_0} u(t,x)= \mathcal{A}(D) u(t,x), \,
t>0, \, x \in \R^n,
\end{equation}
\begin{equation}
\label{innuneg} u(0,x)= \vf (x), \, x \in \R^n.
\end{equation}
\end{theorem}

{\it Proof.} It follows from Theorems \ref{thmweak} and
\ref{thmstrong} that the order function $\beta(\mu t +\nu \tau)$
under the integral in $D_{\ast \{\mu, \nu\}}^{\beta(t)}$ takes the
value $\beta_0$ for all $t < t^{\ast} = min\{t^{\ast}_{1},
T^{\ast}_1\}$ and $0 < \tau < t.$  Hence, the variable order
operator in (\ref{GenEq}) becomes $D_{\ast}^{\beta_0}$ if  $0< t <
t^{\ast}$. The order $\beta_1$ (or diffusion mode
$\{\beta_1,\mu,\nu\}$) has no influence in this interval. For
$t>t^{\ast}$ two diffusion modes $\{\beta_0,\mu,\nu\}$ and
$\{\beta_1,\mu,\nu\}$ are present. If $t < min \{t^{\ast}_2,
T^{\ast}_2\}$ then for all $\tau >0$ we have $\mu t +\nu \tau <T_2$.
That is, there is no influence of the mode $\{\beta_2,\mu,\nu\}$ if
$t < min \{t^{\ast}_2, T^{\ast}_2\}$. In the same manner the other
values of $\beta$ have no influence in the interval $0<t<t^{\ast}.$
This means that for $0<t <t^{\ast}$ solutions of problems
(\ref{GenEq})-(\ref{continuity}) and (\ref{beta0}),(\ref{innuneg})
coincide. \eproof

\section{Some properties of the fundamental solution}

In this section we study some basic properties of a solution
$u(t,x)$ to the problem   (\ref{GenEq})-(\ref{continuity}). Namely,
we show that $u(t,x)$ is a density function under a rather general
conditions on the pseudo-differential operator $\mathcal{A}(D)$. We
also study the MSD(t) of the corresponding process near the initial
time, which is important, in particular, in cell biology.

\begin{theorem}
\label{pdf} Assume $A(\xi)$ is a continuous symbol with negative
values for $\xi \neq 0, \, A(-\xi)=A(\xi)$ and $A(0)=0.$ Then the
Fourier transform $\hat{U}(t,\xi)=F[U](t,\xi)$ of the fundamental
solution $U(t,x)$ to the problem  (\ref{GenEq})-(\ref{continuity})
satisfies the following conditions:
\begin{enumerate}
\item
$\hat{U}(t,\xi)$ is continuous in $\xi$ for every fixed $t \geq 0;$
\item
$\hat{U}(t,0)=1$ for all $t \geq 0;$
\item
$\hat{U}(t,\xi)$ is positive definite for every fixed $t\geq 0.$
\end{enumerate}
\end{theorem}

{\it Proof.} First, let $\nu=0.$ Then Corollary \ref{cor} and the
symmetry $A(-\xi)=A(\xi)$ imply that

\[
\hat{U}(t,\xi)= \mathcal{I}^{'}_0 (t) E_{\beta_0}(t^{\beta_0}
{A}({\xi}))+
\]
\[
\sum_{k=1}^N \mathcal{I}^{'}_k (t)  \, \{
E_{\beta_k}((t-t_{k})^{\beta_k} {A}(\xi))
 \prod_{j=0}^{k-1} E_{\beta_j} ((t_{j+1}-t_{j})^{\beta_j} {A}(\xi)) +
\]
\[
E_{\beta_k} ((t- t_{k})^{\beta_k} A(\xi)) \int_{t_{k-1}}^{t_{k}}
E_{\beta_{k-1}} ((t_{k} - \tau)^{\beta_{k-1}} A(\xi)) \, \, \, \,
_{t_{k-1}}D_{\tau}^{1-\beta_{k-1}} \mathcal{R}_{k-1} (\tau, \xi) d
\tau +
\]
\begin{equation}
\label{ftoffs} \int_{t_{k}}^t E_{\beta_k} ((t- \tau)^{\beta_k}
A(\xi)) \, \, \, \, _{t_{k}}D_{\tau}^{1-\beta_k} \mathcal{R}_k
(\tau, \xi)
 d \tau \, \},
\end{equation}
where $E_{\beta}(z)$ is the Mittag-Leffler function and
$\mathcal{R}_k (t,\xi)$ is defined in (\ref{rk}). Taking into
account continuity of $E_{\beta}(z),$
%alteration
%(in fact the Mittag-Leffler function is entire function for $0<\beta \leq 1$)
we conclude that $\hat{U}(t,\xi)$ is continuous for every fixed $t >
0.$ Further, it follows from the definition of
$\mathcal{R}_k(t,\xi)$ that $\mathcal{R}(t,0)=0.$ This implies
$\hat{U}(t,0)=\sum_{k=0}^N \mathcal{I}_k^{'}(t),$ since
$E_{\beta}(0)=1.$ Finally, as is known \cite{Pollard48,Schneider96},
the function $E_{\beta}(-\lambda t^{\beta}), \, 0<\beta \leq 1,$ is
completely monotone for all $t>0$ if $\lambda >0,$ and therefore
$E_{\beta}(-\lambda t^{\beta}) > 0$ and $d E_{\beta}(-\lambda
t^{\beta})/dt < 0$ for all $t>0.$ The latter implies
$\mathcal{R}_k(t,\xi)\geq 0, \, k=1,...,N.$ It follows from this
fact together with $A(\xi) \leq 0$ and positiveness of
$E_{\beta}(t^{\beta} A(\xi))$ for all $t \geq 0$ that
$\hat{U}(t,\xi) > 0$ for every fixed $t \geq 0$ and $\xi \in K
\subset \subset R^n,$ where $K$ is an arbitrary compact. Now
%applying the Bochner-Schwartz theorem
it is easy to verify positive definiteness of $\hat{U}(t,\xi)$ for
each fixed $t > 0.$ The idea of the proof in the general case $\nu
\in(-1,1]$ is preserved, since the general structure of the
representation formula for the fundamental solution remains
unchanged. \eproof

\begin{corollary}
Under the assumption of Theorem \ref{pdf}  the fundamental solution
$U(t,x)$ to the problem (\ref{GenEq})-(\ref{continuity}) is a
probability density function for each fixed $t \in (0, \infty).$
\end{corollary}

The proof of this statement immediately follows from the
Bochner-Khinchin theorem (see, e.g., \cite{Billingsley}).

Thus there exists a stochastic process $X_t$ with a density function
$p_t (x)=U(t,x)$ for every fixed $t \geq 0$ with $p_0(x) = \delta
(x).$ Denote by $\mu_t=E[X_t]$ the expectation of a random variable
$X_t$ ($t$ is fixed) and $MSD(t)=E[|X_t-\mu_t|^2].$

%alteration{this paragraph is slightly modified}
Now assume that the pseudo-differential operator on the right hand
side of (\ref{GenEq}) is a negative definite second order
homogeneous elliptic operator, that is the symbol of the operator
$\mathcal{A}(D)$ has the form ${A}(\xi) =\frac{1}{2}\sum a_{ij}\xi_i
\xi_j.$ The matrix ${\bf A}= (a_{ij})$ is symmetric and negative
definite: $V^T {\bf A} V \leq - C |V|^2, \, C>0,$ where $V$ is an
$n$-dimensional vector, $V^T$ is its transpose. By $Tr({\bf A})$ we
denote the trace of ${\bf A}$: $Tr({\bf A})= \sum a_{jj}.$ It is not
hard to verify that in this case $U(t, -x)=U(t,x)$ and $\mu_t = 0.$
Hence,
\begin{equation}
\label{msd} MSD(U;t) = \int_{\R^n} |x|^2 U(t,x) dx.
\end{equation}

\begin{theorem}
\label{asymp1} Assume $\mathcal{A}(D)$ in (\ref{GenEq}) is a second
order homogeneous negative definite elliptic operator. Then there
exists $t^{\ast}>0$ such that for $t < t^{\ast}$ the function
$MSD(U;t)$, where  $U(t,x)$ is the fundamental solution of the
Cauchy problem (\ref{GenEq})-(\ref{continuity}), is represented in
the form
\begin{equation}
\label{asnuneg} MSD(U; t) =  \frac{Tr({\bf A})}{\Gamma(\beta + 1)}
t^{\beta}, \,  \, 0< t < t^{\ast}.
\end{equation}
\end{theorem}

{\it Proof.} The proof is an implication of Theorem \ref{nunegative}
and the fact that $MSD(U;t)$ for a solution of the Cauchy problem
(\ref{beta0}),(\ref{innuneg}) can be represented in the form
$MSD(U;t)= (-\Delta_{\xi}) \hat{U}(t, \xi)|_{\xi=0}$ \eproof

%alteration{the following statement added by request of the referee}
\begin{theorem}
Under the condition of Theorem \ref{asymp1} for $MSD(U; t),$ where
$U$ is the fundamental solution of the Cauchy problem
(\ref{GenEq})-(\ref{continuity}), the asymptotic relation
$MSD(U;t)=O(t^{\beta_N}), \, t \rightarrow \infty,$ holds.
\end{theorem}

Proof. It follows from (\ref{ftoffs}) for $t>T_N^{\ast}$ that
$\hat{U(t,\xi)}=Q_1(t,\xi)+Q_{2}(t,\xi),$ where
\[
Q_1(t,\xi)= E_{\beta_N}((t-t_{N})^{\beta_N} {A}(\xi)) ~ \{ ~
\prod_{j=0}^{N-1} E_{\beta_j} ((t_{j+1}-t_{j})^{\beta_j} {A}(\xi)) +
\]
\[
 \int_{t_{N-1}}^{t_{N}} E_{\beta_{N-1}} ((t_{N} -
\tau)^{\beta_{N-1}} A(\xi)) \, \, \, \,
_{t_{N-1}}D_{\tau}^{1-\beta_{N-1}} \mathcal{R}_{N-1} (\tau, \xi) d
\tau ~ \},
\]
and
\[Q_2(t,\xi)= \int_{t_{N}}^t E_{\beta_N} ((t-
\tau)^{\beta_N} A(\xi)) \, \, \, \, _{t_{N}}D_{\tau}^{1-\beta_N}
\mathcal{R}_N (\tau, \xi)
 d \tau.
\]
It is not hard to verify that $Q_1(t,\xi)$ dominates in terms of
asymptotics for large $t.$ We observe the same after applying
$-\Delta_{\xi}$ as well. Elementary, but tedious calculations show
that $(-\Delta_{\xi}) Q_1(t, \xi)=O(t^{\beta_N}), ~ t \rightarrow
\infty.$ \eproof

\begin{corollary}
Let $\beta(t) = \beta$, where $\beta$ is a constant in $(0,1]$. Then
$$ MSD(U; t) =  \frac{Tr({\bf A})}{\Gamma(\beta + 1)} t^{\beta}, \,  \, t > 0. $$
\end{corollary}

%alteration {text of the remark slightly changed with accordance alterations above}
\begin{remark}
A natural generalization of the model is to allow changing of random
diffusion modes $\beta_k$  at random times $T_k$ with appropriate
distributions respectively. The questions on an asymptotic behaviour
of $U(t,x)$ for large times, which shows how heavy is the tail of
the distribution, as well as an asymptotic behaviour of $MSD(U;t),
\, t \rightarrow \infty,$ which tells about the nature of the
corresponding process, are important. We will discuss these
challenging questions in a separate paper.
\end{remark}

%alteration
{\bf {\em Acknowledgment.}} We are thankful to Professor C. Lorenzo
for his useful comments. We also acknowledge thoughtful remarks by
anonymous referees, which essentially improved the text. The
research is supported by NIH Grant P20 GMO67594.

\end{document}